\documentclass[prx,reprint,flushbottom,amsmath,amssymb,longbibliography]{revtex4-1}
\usepackage{graphicx,epsfig,epsf,color,hhline}
\usepackage{dcolumn}
\usepackage{bm}
\usepackage{tensor,pbox}

\newcommand{\ket}[1]{\left|#1\right>}
\newcommand{\bra}[1]{\left<#1\right|}
\def\ketbra#1{|#1\rangle\langle#1|}
\newcommand{\nn}{\nonumber\\}

\newcommand{\bea}{\begin{eqnarray}}
\newcommand{\ea}{\end{eqnarray}}
\newcommand{\eea}{\end{eqnarray}}
\newcommand{\ord}{{\cal O}}
\newcommand{\sumint}[1]

\begin{document}
\newcommand{\ri}{ i}
\newcommand{\re}{ e}
\newcommand{\bx}{{\bm x}}
\newcommand{\bd}{{\bm d}}
\newcommand{\be}{{\bm e}}
\newcommand{\br}{{\bm r}}
\newcommand{\bk}{{\bm k}}
\newcommand{\bA}{{\bm A}}
\newcommand{\bE}{{\bm E}}
\newcommand{\bB}{{\bm B}}
\newcommand{\bI}{{\bm I}}
\newcommand{\bH}{{\bm H}}
\newcommand{\bR}{{\bm R}}
\newcommand{\bZero}{{\bm 0}}
\newcommand{\bM}{{\bm M}}
\newcommand{\bX}{{\bm X}}
\newcommand{\bn}{{\bm n}}
\newcommand{\bs}{{\bm s}}
\newcommand{\bv}{{\bm v}}
\newcommand{\tbs}{\tilde{\bm s}}
\newcommand{\rSi}{{\rm Si}}
\newcommand{\beps}{\mbox{\boldmath{$\epsilon$}}}
\newcommand{\bGamma}{\mbox{\boldmath{$\Gamma$}}}
\newcommand{\bxi}{\mbox{\boldmath{$\xi$}}}
\newcommand{\rg}{{\rm g}}
\newcommand{\tr}{{\rm tr}}
\newcommand{\xmax}{x_{\rm max}}
\newcommand{\xb}{\overline{x}}
\newcommand{\pb}{\overline{p}}
\newcommand{\ra}{{\rm a}}
\newcommand{\rx}{{\rm x}}
\newcommand{\rs}{{\rm s}}
\newcommand{\rP}{{\rm P}}
\newcommand{\up}{\uparrow}
\newcommand{\down}{\downarrow}
\newcommand{\hc}{H_{\rm cond}}
\newcommand{\kb}{k_{\rm B}}
\newcommand{\cI}{{\cal I}}
\newcommand{\tit}{\tilde{t}}
\newcommand{\cE}{{\cal E}}
\newcommand{\cC}{{\cal C}}
\newcommand{\Ubs}{U_{\rm BS}}
\newcommand{\sech}{{\rm sech}}
\newcommand{\qq}{{\bf ???}}
\newcommand*{\etal}{\textit{et al.}}
\def\vec#1{\bm{#1}}
\def\ket#1{|#1\rangle}
\def\bra#1{\langle#1|}
\def\keps{\bm{k}\boldsymbol{\varepsilon}}
\def\dm{\boldsymbol{\wp}}

\title{Intrinsic measurement errors for the speed of light in vacuum}
  
\author{Daniel Braun$^1$, Fabienne Schneiter$^1$, and Uwe R. Fischer$^2$}
\affiliation{$^1$Eberhard-Karls-Universit\"at T\"ubingen, Institut f\"ur Theoretische Physik, 72076 T\"ubingen, Germany\\
$^2$Seoul National University, Department of Physics and Astronomy \\  Center for Theoretical Physics, 08826 Seoul, Korea}

\begin{abstract}
The speed of light in vacuum, one of the most important and
 precisely measured natural constants,
is fixed by convention to $c=299 792 458$ m/s. Advanced theories 
predict possible deviations from this universal value, or even quantum
fluctuations of $c$.  Combining arguments from quantum parameter estimation
theory and classical general relativity, we here establish rigorously the
existence of lower bounds on the uncertainty to which the speed of light
in vacuum can be determined in a given region of space-time, {subject
to several reasonable restrictions.}
They provide a novel perspective on the experimental falsifiability of predictions for the quantum fluctuations of space-time.
 
\end{abstract}
\maketitle

\section{Introduction}
It is generally accepted that the speed of light in vacuum $c$ is a
universal natural constant, isotropic, independent of frequency, and
independent 
of the motion of the inertial frame with respect to which it is
measured.  These properties have been experimentally demonstrated with
very high precision, e.g.~isotropy up to a
relative uncertainty of the order of $\sim 10^{-9}$
\cite{Scheithauer05},
and lie at the basis
of special relativity.  
By 1972, measurements of the speed of light  
became more precise 
than the definition of the meter \cite{evenson_speed_1972}, leading
in 1983 to the definition of the speed of light in vacuum $c=299
792 458$\,m/s.  
But attempts to quantize gravity have led to the concept of 
space-time as a fuzzy ``quantum foam'' on the
Planck length $l_{\rm Pl}=\sqrt{\hbar G/c^3}\simeq 1.62 \times 10^{-35}\,$m  
{\cite{snyder_quantized_1947,doplicher_quantum_1995,QuantumFoam}} 
that implies an uncertainty or dispersion of $c$ 
\cite{Hawking80,Ashtekar92,Ford95,Yu00}.
Experimental data based on gamma-ray bursts, pulsars,  and TeV-flares
from active galaxies imply upper 
bounds on deviations of $c$ 
over cosmic distances 
{\cite{Amelino-Camelia98,Ellis00,Biller99,Kaaret99,Amelino-Camelia14,vasileiou_planck-scale_2015,Perlman}}.   
Quantum fluctuations of $c$ were also proposed
due to virtual fermion-anti-fermion pairs, 
leading to 
a scaling of the jitter of the arrival time of light pulses with
propagation distance \cite{Urban13,Leuchs13}. 
Satellite experiments are being planned to verify fundamental
space-time properties with unprecedented precision, such as the isotropy of $c$ and its  independence from the laboratory frame velocity \cite{Scheithauer05}. 

Here we establish how
precisely $c$ in a given region of space--time may be 
determined {\em in principle}, i.e.~independent of any technical
challenges.  Our approach is based
on the firmly 
established quantum parameter estimation theory (q-pet)
\cite{Helstrom1967,helstrom_quantum_1969,Braunstein94,braunstein_generalized_1996,wiseman_quantum_2009,Giovannetti04,giovannetti_quantum_2006,Paris09}
and general relativity (GR) in semiclassical approximation {\cite{davies_birrell}}.  
Q-pet allows one to obtain a lower bound
on the uncertainty with which a parameter $\theta$ 
may be estimated that
parametrizes a quantum state specified by  a density matrix $\rho(\theta)$. The
power of q-pet is due to the facts that {\em i.)} the bound is reachable
in the limit of a large number of measurements, and {\em ii.)} it is
optimized over all possible  
quantum mechanical measurements (positive operator valued measures,
POVM \cite{Peres93}) and all data-analysis schemes (unbiased estimator
functions). This so-called quantum Cram\'er-Rao bound (QCRB)
\cite{Helstrom1967,helstrom_quantum_1969,Braunstein94,braunstein_generalized_1996} 
becomes relevant once all technical
noise problems have been solved, and only the fundamental quantum
uncertainties remain.  It is the ultimate achievable lower bound on
the uncertainty with which any parameter can be measured.
Recently, the q-pet formalism was applied to the
measurement of parameters in relativistic quantum field
theory such as 
proper times and accelerations, the Unruh effect, gravitation, or the
estimation of 
the mass of a black hole \cite{Ahmadi14,Ahmadi14.2,Aspachs10,Downes11}.  In the
present work we go a step further by examining the back-action of the
quantum probe on the metric of space-time. Taking back-action into account
was proposed before
\cite{RevModPhys.29.255,PhysRev.109.571,Ahluwalia94,Amelino-Camelia94,Ng00}
but to the best of our knowledge we 
combine for the first 
time modern q-pet 
with a precise calculation of the back-action of the probe on the space-time
metric. 
We show that there is an optimal photon number at which the perturbation of
the space-time metric due to the probe equals the quantum uncertainty
of the measurement itself, establishing thus an ultimate lower bound
on the uncertainty with which $c$ can be determined.  

\section{Quantum parameter estimation}
Any {\em direct} measurement of the speed of light has to use a
light signal. Indirect measurements,
e.g.~through measuring the fine-structure constant, the electron charge and
Planck's constant, may need no light but do
not reflect the definition of $c$ as a speed
and need an 
elaborate theoretical framework for their interpretation. 
{We consider {definitions} of $c$ through $c=\Delta x/\Delta t$
(i.e.~runtime measurements of a light pulse) as well as
through $c=\omega/k$ (where $\omega$ is (2$\pi$ times) the frequency
and $k$ the wavevector of a monochromatic e.m.~wave) as direct
measurements, as these {\em i.)} use a light signal; {\em ii.)}
correspond to how $c$ has actually 
been determined experimentally (in particular the most precise
determinations of $c$ 
to date use $c=\omega/k$ \cite{evenson_speed_1972}), and {\em iii.)}
are based on simple 
three-letter formulas that need no elaborate theoretical framework for
extracting $c$. These two definitions give $c$ the meaning of a
propagation speed or phase speed, respectively. Note that we only need
$c=\omega/k$ at the frequency considered, not over all
frequencies. For wave-lengths comparable to quantum-gravity length
scales (assumed to be of order Planck-length), modifications of this
linear dispersion relation have been proposed {(see the discussion on
rainbow gravity in Sec.\ref{sec.rainbow})}, but we restrict 
ourselves to frequencies where the linear dispersion is well verified
experimentally.} { We emphasize that these definitions of speed are
only needed to determine a systematic experimental error due to GR
effects. The quantum-mechanical uncertainty of $c$ obtained from q-pet
on the other hand is optimized over all possible (POVM) measurements of the
light signal and analysis schemees of the data, including those that measure
the propagation distance $\Delta x$ of a light pulse over a
time-interval $\Delta t$. We therefore do not have to worry about
additional uncertainties of measurements of positions or times.}

Any light signal
can be decomposed in modes of the electromagnetic (e.m.)~field
which are the fundamental dynamical objects  in quantum optics. 
Q-pet shows that with $m$ modes the
sensitivity can be improved at most by a factor $1/m$
\cite{giovannetti_quantum_2006}. Below we 
find that with at most $n$ photons in a single mode the best sensitivity
scales as $\propto 1/n$; one can thus achieve for given maximum photon
number $nm$  the same  
sensitivity scaling as $\propto 1/(nm)$  as with $m$ modes (for a
strict proof see Appendix \ref{singlemode}). In
\cite{giovannetti_quantum-enhanced_2001} the  
problems of positioning and clock synchronization were
analyzed.  They were reduced to measuring a travel time of a light
pulse with constant $c$, which is closely related to 
measuring $c$ for a known propagation distance.  Also there it was shown that the
best uncertainty in the arrival time of the pulse for a squeezed m-mode
state scales as $1/(nm)$. Furthermore, using the Margolus-Levitin
quantum speed limit theorem, it was argued in
\cite{giovannetti_quantum-enhanced_2001} that this is the optimal 
scaling possible for any state.
The scaling $\propto 1/\bar{n}$ for large average photon number
$\bar{n}$ was also obtained for phase estimation with two-mode 
squeezed light in \cite{anisimov_quantum_2010}.
As for relativistic effects, if we are 
interested in knowing $c$ in a given space-time region, they cannot be
diluted by using several modes in 
parallel in different space-regions or sequentially. 
We can thus restrict ourselves to studying a single mode.
For concreteness, we consider a cubic cavity with edges
of length $L$,  
and perfectly reflecting walls or symmetric boundary conditions.  

Maxwell's equations in vacuum with appropriate boundary conditions impose quantized
modes with wave vectors $\bk$ that are independent of 
$c$, whereas the frequency $\omega=c|\bk|$.
Obtaining the best 
possible precision of $c$ 
is thus equivalent to the optimal frequency measurement of a harmonic oscillator, for which
the quantum Cram\'er-Rao bound was calculated
in \cite{Braun11.2}.
The smallest $\delta \omega/\omega$, and
hence 
smallest $\delta c/c$ for fixed
maximum excitation $2n$ and for $\tau=\omega t\gg 1$, is achieved with
the optimal state
$|\psi_{\rm opt}\rangle=(|0\rangle+|2n\rangle)/\sqrt{2}$. In a single
measurement, it leads to a minimal 
$c$-uncertainty 
\begin{equation}
  \label{eq:opt}
  \frac{\delta
    c}{c}\simeq \frac{1}{2\tau n}.
\end{equation}

For existing measurements with large $n$, coherent states are  
more relevant than the optimal state. A coherent state 
with amplitude 
$\alpha$ at time $t=0$, $|\psi_{\rm coh}\rangle =|\alpha\rangle$,
evolves according to
$\alpha(t)=\alpha e^{-i\omega t}$  \cite{Scully97} and leads to 
\begin{equation}\label{eq:dccoh}
  \frac{\delta
    c}{c}=\frac{1}{2}\frac{1}{|(\frac{1}{2}+n)\sin^2\tau+n\tau(\tau+\sin(2\tau))|^{1/2}}
  \simeq  \frac{1}{2\tau\sqrt{n}},
\end{equation}
where the last equality is for large $\tau=\omega t$ and large average
photon number $n=\alpha^2$ ($\tau^2\alpha\gg 1$) \cite{Braun11.2}. 

From these results one is tempted to conclude that ${\delta c}/{c}$ can be made 
arbitrarily small by increasing $n$. 
However, the 
energy-momentum tensor
increases $\propto n$ for $n\gg 1$, and will at some point perturb itself
the metric of space-time.  
We argue that the ultimate sensitivity is reached when the general
relativistic modification
of 
space-time 
becomes comparable to the minimal quantum uncertainty of the measurement.
This leads to a finite optimal number of photons, and a finite optimal
sensitivity. Increasing the photon number even more will modify
space-time 
to a point where one cannot speak of light
propagation in vacuum anymore. In principle one may re-calculate from
the measured value using GR what
the speed of light in vacuum {\em would} be, but this is a counterfactual
reasoning and not a direct measurement of $c$. On the other 
hand, 
reducing the photon number would 
increase the quantum noise. The situation is very similar to the
optimization of the photon number in LIGO-type gravitational wave
interferometers, where one balances photon-shot noise against
radiation pressure noise
\cite{Caves81,abadie_gravitational_2011,demkowicz-dobrzanski_fundamental_2013}.
However,  
whereas radiation pressure noise is 
specific to the measurement instrument, 
in our case the properties of space-time itself and thus the very
meaning of light propagation in vacuum are 
affected when increasing the photon number further, and 
this effect is unavoidable.

The gravitational effects sought here are
well in the 
regime where Einstein's field equations are valid: 
Firstly, we consider light at
wavelengths $\lambda$ and structures of the energy-momentum tensor on
scales much larger than the Planck-length (e.g.~$\lambda=500$\,nm and a
standard (possibly lossy) cavity of size $L=1$\,km).  
Secondly, we consider light fields 
of very large intensity and effects linear in the perturbation of the metric, for which the
energy-momentum tensor should be  
well approximated by its quantum mechanical
expectation value \cite{FORD1982238}. 
It is the effect
of this average energy-momentum tensor on space-time that we calculate 
and compare to the minimal uncertainty of $c$ obtained from q-pet, not the
fluctuations of space-time themselves.
The former is
established on the solid ground of general relativity, 
whereas the latter would require a quantum gravity theory to make
reliable predictions. The  
quantum fluctuations that we {\em are} interested in here are those of
light probing the space-time, which are reliably described by quantum optics. 
Our results therefore rely only on well-tested theories, 
in distinction to predictions of the fluctuations of space-time
obtained by various theories of quantum gravity.

\section{Perturbation of metric due to light intensity} 
The modification of the metric of
space-time is found from the weak  
field limit of the Einstein field equations, 
where  the metric tensor is given by
$g_{\mu\nu}=\eta_{\mu\nu}+h_{\mu\nu}$, i.e.~the flat Minkowski metric 
$\eta_{\mu\nu}={\rm
  diag}(-1,1,1,1)$ (in terms of $ct,x,y,z$)
plus a small perturbation, 
$|h_{\mu\nu}|\ll 1$.
Einstein's  equations yield a wave
equation for the trace inverse, $\bar{h}^{\mu\nu}=h^{\mu\nu}-\frac{1}{2}\eta^{\mu\nu}\eta^{\alpha\beta}
h_{\beta\alpha}$, 
\begin{equation}
  \label{eq:weakfield}
\square \bar{h}^{\mu\nu}=-16\pi \frac{G}{c^4} T^{\mu\nu} ,
\end{equation}
where the (flat space-time) Lorenz gauge (FLG) condition
$\bar h^{\mu\nu}{}_{,\nu}=0$ is used; see eq.~18.8b in
\cite{MTW}. 
The energy-momentum tensor $T^{\mu\nu}$ of the e.m.~field reads \cite{MTW}
\begin{eqnarray}
  \label{eq:tmn}
  T^{00}&=&\frac{1}{2}(\epsilon_0\bE^2+\mu_0\bH^2), \quad 
T^{0i}=T^{i0}=\frac{1}{c}(\bE\times \bH)_i,\nn
T^{ij}&=&-\left(\epsilon_0E_iE_j+\mu_0H_iH_j\right)+T^{00}\delta_{ij},
\end{eqnarray}
where $i,j\in\{1,2,3\}=\{x,y,z\}$. We use the
q.m.~expectation value of $T^{\mu\nu}$ as source term in
(\ref{eq:weakfield}) for the (011) and the (01M) modes
($k_i=l_i\pi/L$, $l_x=0,l_y=1$, and $l_z=1$ or $l_z=M$, respectively;
$\Omega_l=c|\bk|$, and $V=L^3$). This ``semiclassical approximation''
is justified if one is interested only in effects to first order in
$h_{\mu\nu}$ \cite{FORD1982238}. {Using the (011) mode is
  motivated by the fact that it has lowest frequency and hence
  expected lowest GR impact. This will be verified by comparing to the
  (01M) mode with large $M$.} For $\ket{\psi_{\rm
    opt}}$ with $n\gg 1$, the  
solution of (\ref{eq:weakfield}) for the (011) mode reads  ($\bxi=\pi\bx/L$)
\begin{eqnarray}
  \bar h^{\mu\nu}(\bxi)&=& {\mathcal P} \int_0^\pi\int_0^\pi
d\eta'd\zeta'I(\xi,\eta-\eta',\zeta-\zeta')t^{\mu\nu}(\eta',\zeta'),\nonumber\\  
  \label{eq:A}
  {\mathcal P}&=&\frac{4\sqrt{2}n}{\pi}\kappa ,\quad\quad
  \kappa=\left(\frac{l_{\rm 
        Pl}}{L}\right)^2,\\ 
  I(\xi,\eta,\zeta)&=&\ln\left(\frac{\xi+\sqrt{\xi^2+\eta^2+\zeta^2} 
    }{\xi-\pi+\sqrt{(\xi-\pi)^2+\eta^2+\zeta^2
    }}\right),
\end{eqnarray}
with 
dimensionless trigonometric functions $t^{\mu\nu}:=
T^{\mu\nu}/(n\hbar\Omega_l/V)$ of order one inside the cavity, 
and zero outside (see Appendix \ref{appendixmetric}). 
$T\indices{^\mu_\mu}=0$ for the
e.m.~field \cite{LandauLifschitz87}, hence $h\indices{^\mu_\mu}=0$ and 
$h^{\mu\nu}=\bar{h}^{\mu\nu}$. 

The deviations of $\bar h^{\mu\nu}$ in (\ref{eq:A})
from FLG are of second order
in $h$ and can be neglected \cite{Eddington}.  
For $\ket{\psi_{\rm coh}}$, $\bar{h}^{\mu\nu}$ is
the same as for $\ket{\psi_{\rm opt}}$
plus retarded oscillation on top of it, with an amplitude of the same
order.  We therefore restrict the analysis to the
time-independent part. For the $(01M)$ 
mode, and $n,M\gg 1$, only ${h}^{00}$
and ${h}^{33}$ are non-negligible, 
\begin{eqnarray}
  \label{eq:h00}
 {h}^{00}={h}^{33}\simeq 4{\mathcal P} M \int_0^\pi\int_0^\pi \!
d\eta'd\zeta'I(\xi,\eta-\eta',\zeta-\zeta')\sin^2\eta' .\nonumber
\end{eqnarray}
From the geodesic 
condition $ds^2=g_{\mu\nu}dx^\mu dx^\nu=0$, the local modification of the coordinate speed of
light 
\begin{eqnarray}
  \label{eq:dc2}
  \delta c(x)/c = -\frac12 (h_{00} + h_{11})
\end{eqnarray}
 is obtained for the (011) mode, with similar expressions for $\delta
 c(y)$ and $\delta c(z)$ (see also Fig.\,\ref{fig.dc} in Appendix \ref{appendixmetric}). 
For the $(01M)$ mode with $n,M\gg 1$, 
$
\delta c(x)/c =\delta c(y)/c =-\frac12 h_{00},\,\,\delta c(z)/c=2  \delta c(x)/c.
$
One may object that according to the
equivalence principle one could always find a
coordinate system (CS) in which $c(x)=c(y)=c(z)=c$, and that
by the definition of $c$ one {\em should} go to the free falling
CS for measuring $c$, where $c$ is always the same.  However, one has
to distinguish between the 
universal constant $c$ {entering Lorentz-transformations}, and the experimental value $c_{\rm exp}$
{of the propagation speed of light} obtained in measurements.  The
experimental definition of $c$,  $c_{\rm exp}=\Delta x/ \Delta t$,
where $\Delta x$ is the 
distance that a light signal travels in time $\Delta t$ 
implies that for any finite $\Delta x$ 
the measurement is non-local, which precludes transforming the
discussed GR effect away by a local transformation. It is to
be expected that this non-local effect can be made arbitrarily small
by moving the two points arbitrarily close to each other. More
importantly, however, the {measurement apparatus}
cannot be free falling in the gravitational field of the light
it contains, as it carries that light with it. A time delay can be
measured with a single clock by passing {a short light pulse}
through a beam 
splitter (BS), reflecting it on a mirror and sending it back to the
BS. The two passes through the BS trigger start/stop of the clock {by
light scattered into detectors adjacent to the BS}. The
clock measures its proper time, $d \tau=  
\sqrt{-g_{00}} dt$. {$\Delta x$ has to be measured independently,
i.e.~with standard measurement rods. 
  Hence,
$\Delta x$ corresponds to the ``proper length''  of the
apparatus (distance between BS and mirror for a runtime experiment or
length of the cavity when using $\omega=c k$).  ``Proper length'' (not
to be confused with ``proper 
distance'')  is defined as
the length measured with standard measurement rods in the frame where
the object is at rest \cite{Fayngold09}.  We may assume the measurement
rods as well as the 
measurement apparatus as sufficiently ``rigid'' (gravitational forces
and modification of the e.m.~forces that 
determine the shapes of these objects  much
smaller than the e.m.~forces 
that determine their shape and arrangement
\cite{grishchuk_emission_1974,kopeikin_optical_2015}), which 
means that $\Delta x$ remains
unchanged when the light intensity is increased.} In the limit
$R\gg L$ ($R$ = typical radius of 
curvature of space time), the
experimentally found value $c_{\rm 
  exp}(x)= \Delta x/\Delta \tau\simeq
dx/d\tau=c(x)/\sqrt{-g_{00}}$ is then directly related to the
coordinate speed $c(x)$ determined above.  This gives $\delta c_{\rm
  exp}/c=-h_{11}/2$ for the (011) mode, where $\delta c_{\rm
  exp}(x):=  c_{\rm  exp}(x)-c $ can be
 considered a systematic error in the determination of $c$. 

Since q-pet was based on the uncertainties 
$\delta\omega$, we also
compare q-pet and GR based on the GR shift of the
cavity resonance frequencies
by solving the e.m.~wave
equation in the entire cavity with mirrors at 
$0,x_L$ and symmetric boundary conditions (SBC),  
$A^\mu(0,y,z)=A^\mu(x_L,y,z)$ (and correspondingly for the other
directions).  The unperturbed single modes are plane waves  
 $A^3(t,x,y,z)=\left({\hbar}/{(2\omega\epsilon_0
     V)}\right)^{1/2}
     (e^{ik(x-ct)}a+{\rm h.c.})$,
$A^\mu=0$ for $\mu\in\{0,1,2\}$, and $k:=k_0= k_1>0$. This leads to
$T^{\mu\nu}=-\hbar\omega/(2\epsilon_0V)\langle
(a\,e^{ik(x-ct)}+{\rm h.c.})^2\rangle$ for
$(\mu,\nu)\in\{00,01,10,11\}$ inside the cavity, and $T^{\mu\nu}=0$ else or
outside. For $\ket{\psi_{\rm opt}}$,
$T^{\mu\nu}$ is time-independent, and for $\ket{\psi_{\rm coh}}$ we once
more consider only the time-independent part. Then, 
$h^{\mu\nu}(\bxi)=\epsilon(\bxi)$ for 
$(\mu,\nu)\in\{00,01,10,11\}$ and $h^{\mu\nu}=0$ else, where
\begin{equation}
  \label{eq:fx}
  \epsilon(\bxi) :=\sqrt{2} {\mathcal
  P}M\int_0^\pi\int_0^\pi d\eta'd\zeta'I(\xi,\eta-\eta',\zeta-\zeta').
\end{equation}
The wave equation describing the propagation of light in curved
space-time reads
$
  \nabla_\beta F^{\alpha\beta}=0
$
(see 22.17a in \cite{MTW}),  
with $F^{\alpha\beta}=g^{\alpha\mu}g^{\beta\nu}(A_{\nu,\mu}-A_{\mu,\nu})$.
Using FLG for $A$ and $h$, and $h\indices{^\nu_\nu}=0$, we obtain to first order 
in  $\epsilon$ 
 \begin{eqnarray}
   \label{eq:Awavefin}
  0=-A\indices{^{\alpha,\nu}_\nu}+(h\indices{^{\alpha}_{\mu,\nu}}-h\indices{^\alpha_{\nu,\mu}})A^{\mu,\nu} ,
 \end{eqnarray}
where indices are pulled up or down by the full metric $g^{\mu\nu}$.
Eq.\,\eqref{eq:Awavefin} is solved exactly by the original plane wave
despite the changed metric, as the $\epsilon$-correction in $A^3$ is $\propto
(\partial_x+\partial_{ct})^2$. This reflects the well-known result
that two parallely propagating beams of light do not affect each other
gravitationally
\cite{tolman_gravitational_1931,bonnor_gravitational_1969}. 
The existence of a mode with unchanged dispersion relation suggests
that judging whether the vacuum may still be 
considered as such based on the change of a single mode frequency can
be insufficient. In such a case, the change of the metric can normally
still be probed using other modes. In the example 
above the frequencies of modes propagating in different directions,
e.g.~$A^3\propto \exp(i k(x+ct))$, are modified locally  by a 
relative amount of order $\epsilon(\bx)$, as can be shown by solving
\eqref{eq:Awavefin} in eikonal approximation. 

To summarize, up to numerical prefactors of
  order 1, both systematic { errors $\delta c_{\rm exp}$ obtained by
  measuring length over time  
or a shift } of a cavity resonance, possibly in another mode,
scale as                   
\begin{equation}
  \label{eq:dcGR}
  \frac{\delta c_{{\rm exp}}}{c}\sim 
  -\kappa\,n\,M\,
\end{equation}
for $n,M\gg 1$. With this, we can now obtain the
smallest possible uncertainty with which $c$ can be determined
{in a given region of space-time}. 

\section{Minimal uncertainty of speed-of-light measurements}
For $\ket{\psi_{\rm opt}}$, equating (\ref{eq:opt}) and the absolute value of (\ref{eq:dcGR}) 
leads with $M\sim L/\lambda$ 
to an optimal photon number $n_{\rm opt}\sim (\lambda/l_{\rm
  Pl}) \sqrt{{L}/{(c T)}}$, and a minimal 
\begin{equation}
  \label{eq:dcopt}
  \frac{\delta c}{c}\sim\frac{l_{\rm
      Pl}}{(c\,T\,L)^{1/2}}, 
\end{equation}
{\em independent of frequency}: the gain in quantum
mechanical sensitivity due to longer dimensionless evolution time for
more energetic photons is exactly cancelled by the increased
perturbation of the metric. 

In an experiment, the measurement time is
bounded {from above} by the finite photon-storage time of the
photons in the 
cavity. 
 While obtaining optimal bounds including photon loss requires mixed
 state q-pet \cite{escher_general_2011,Kolodynski10}, 
 the sensitivity cannot be better than that obtained from the pure
 states from which the state is mixed \cite{Braun10}. {For known
   dissipation and decoherence mechanisms one can try to find an
   adapted optimal state.  However, the sensitivity cannot be better
   than if one had access to the full system and its environment. For
   photon loss the environment can be modelled by additional modes
   coupled to the central mode by 
   beam-splitter couplings, and including such ancilla modes cannot
   improve the estimation of a parameter of the original system when
   optimized over all initial states
   \cite{fraisse_hamiltonian_2016,boixo_generalized_2007}, if the
   ancillas are independent of the $c$ we are interested in (which is
   the case for the modes outside the cavity and hence outside the
   space-time region considered).  Our q-pet bound
   calculated for the ideal situation without photon loss therefore
   remains valid, but can in general in the presence of dissipation or
   decoherence not be reached anymore.}  
For a cavity of length $L$ and finesse $F$, 
{the measurement time is bounded by} $T=L
F/(\pi c)$. 
This leads to an optimal number
of photons {\em independent of the length of the cavity}, $n\sim
\lambda/(l_{\rm Pl} F^{1/2})$. For numerical estimates we use in the
following a standard situation: visible light with $\lambda=500$\,nm, a
finesse $F=10000$, and $L=1000$\,m.  The optimal $n$ for the optimal 
state is then 
$n\sim 10^{26}$, and the minimal uncertainty 
$\delta c/c 
\sim l_{\rm Pl}/(L F^{1/2}) \sim 10^{-40}$. 

\begin{figure}[t]
\centering
\includegraphics[width=\columnwidth]{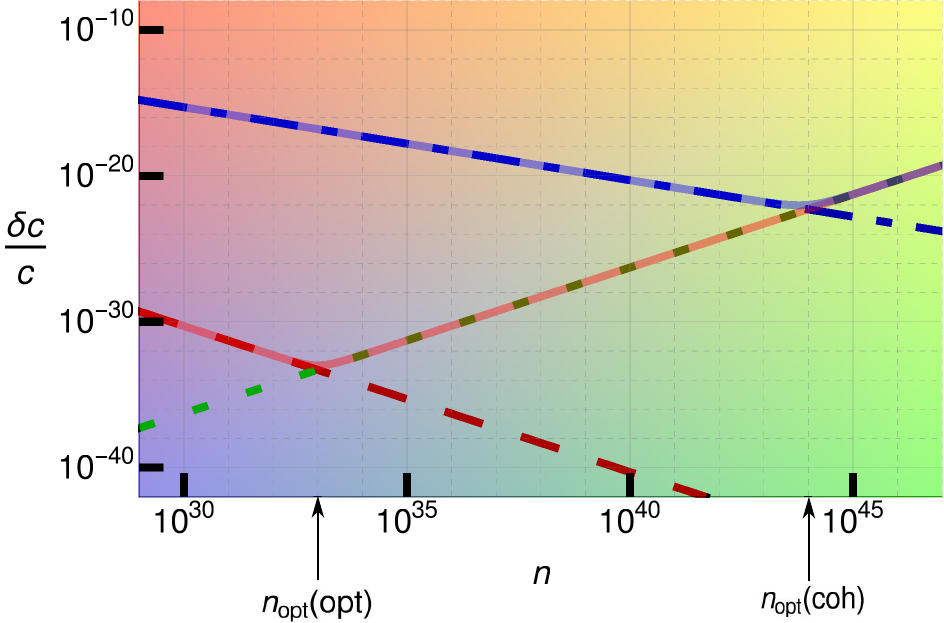}
\caption{Minimal uncertainty ${\delta c}/{c}$ as a function of the
  number of photons $n$: The dashed red/blue line shows the minimal
  uncertainty obtained from the quantum Cram\'er-Rao bound for the
  optimal and coherent states  
given in equations \eqref{eq:opt} and \eqref{eq:dccoh}, respectively. The dashed green line corresponds 
to the unavoidable systematic error in the measurement of $c$ due to
the light's own gravitational effect.
The sum of the minimal uncertainty given by the quantum Cram\'er-Rao bound and the systematic error for the optimal/coherent state is shown by the solid orange/light blue lines. The optimal number of photons minimizing ${\delta c}/{c}$ for either optimal 
or coherent states lies at the minima of the solid orange/light blue lines.
Parameters are $\lambda=500\, {\rm nm}$, $\tau=1$, $L=1\,{\rm km}$ und
$M={L}/{\lambda}$.
}
\label{figscalings}
\end{figure}

For $\ket{\psi_{\rm coh}}$, equating 
(\ref{eq:dccoh}) and (\ref{eq:dcGR}) 
leads to 
$n_{\rm opt}
\sim(L\lambda^2/(l_{\rm Pl}^2 c T))^{2/3}$, and a minimal uncertainty 
\begin{equation}
  \label{eq:dcoptF1}
  \frac{\delta c}{c}\sim
  \left(\frac{l_{\rm Pl}^2\lambda }{L(c\,T)^2}\right)^{1/3}. 
\end{equation}
For a cavity with finesse $F$, the length of the cavity is again 
irrelevant for the optimal photon number, $n_{\rm opt}\sim
\left({\lambda}/{l_{\rm Pl}}\right)^{4/3}/F^{2/3}$, and
$  {\delta c}/{c}\sim {l_{\rm Pl}^{2/3}\lambda^{1/3}}/({LF^{2/3}}). 
      $
Contrary to $\ket{\psi_{\rm opt}}$, the minimal uncertainty
depends here on the wavelength.  In principle, $\delta c/c$ could
therefore be
smaller for $\ket{\psi_{\rm coh}}$ than for
$\ket{\psi_{\rm opt}}$, but only for wavelengths
$\lambda<l_{\rm Pl}\sqrt{c T/L}$ in lossless cavities, and for  
$\lambda< l_{\rm Pl} \sqrt{F}$ in cavities with finesse $F$, 
which are outside the validity of the theory.
For the lossy cavity considered, the optimal coherent state photon
number 
is $n\sim 10^{35}$ and $\delta c/c\gtrsim
10^{-31}$, demonstrating the superiority of $\ket{\psi_{\rm opt}}$.  
  We display the various {$n$}-scaling regimes 
  and the optimal photon numbers located at the minima of the overall dependence
  of $\delta c/c$  on {$n$} in Fig.\,\ref{figscalings}.

{\section{Comparison with similar bounds}
The minimal uncertainties of $c$ and hence the metric of flat
space-time that we have derived   
are reminiscent of ideas about the fuzziness of space-time on the
Planck scale, their different physical meaning not withstanding. The minimal
uncertainty of $\delta c$ that we have derived here translates, in
experiments where a length $L$ is measured through $L=cT$, to
fluctuations $\delta L$ of $L$. There
has been a vast amount of work aiming at demonstrating a minimal
length scale in physics and working out its consequences, see
\cite{hossenfelder_minimal_2013} for an excellent review. The majority of these
works has tried to establish smallest uncertainties of positions or
length measurements, but there have also been attempts to find
minimal uncertainties of volumes, areas,  gravitational fields, event
horizons, and 
others. Here we focus on previous predictions of minimal 
uncertainties of lengths or positions.  For simplicity we set
$\hbar=c=1$ in the rest of this
section and neglect factors of order 1, unless otherwise noted.  }

\subsection{Previous thought experiments}\label{sec.TE}
{
Closest to our analysis are previous thought experiments that one way
or another use classical gravity effects to bound quantum
uncertainties from below.  An illustrative example is the Heisenberg
microscope with gravity \cite{PhysRev.135.B849}. In addition to the
familiar Heisenberg 
microscope, where attempts to resolve the position of a particle by
scattering light from it result in an
unknown momentum kick of order $\omega$ onto the particle, while
limiting the spatial resolution to roughly the 
wavelength of the light $\delta x_{QM}\sim 1/\omega$, one also
considers the gravitational interaction of the photon with the
particle.  This leads to an acceleration of
the particle of at least $~G\omega/R^2$  if the photon is detected at
distance $R$, and a corresponding displacement between the
photon-particle interaction and the photon detection of order
$\delta x_{GR}\sim G\omega$. Taking the geometric mean of the two
uncertainties gives immediately $\delta x\sim
\sqrt{G}=l_\text{Pl}$. Alternatively, we can take the sum of the two
uncertainties and minimize it over $\omega$.  This gives $\omega\sim
1/\sqrt{G}=m_\text{Pl}$, the Planck mass, and, up to a factor 2, again
$\delta x\sim l_\text{Pl}$.  }

{
Another popular argument goes back
at least to Bronstein in 1936 \cite{bronstein_quantentheorie_1936}, who, in the context of
investigating how precisely a gravitational field might possibly be
measured, came up with the request that the test particle should not
collapse to a black hole. Later,  Wigner and Salecker introduced 
a similar limitation to length measurements with light pulses
\cite{RevModPhys.29.255,PhysRev.109.571}, where the clock should not become a
black hole. The idea was refined for the measurement of lengths based
on ``material reference systems'' (MRS)
\cite{Amelino-Camelia94}, consisting of reference
points of size $s$ and mass $M$ that contain a clock,
light-gun and detector, arranged in space. The request that no
event-horizon should form around the reference points beyond $s$
implies $M<s/l_\text{Pl}^2$. }

{We can apply the black-hole  argument to the
Heisenberg-microscope, requesting that the photon's event horizon
should be at least smaller than the distance $R$,
i.e.~$\omega<R/l_\text{Pl}^2$. Then $\delta x_\text{QM}\gtrsim
l_\text{Pl}^2/R$, a bound obviously much weaker than the
previous one for $R\gg l_\text{Pl}$. On the other hand, for the MRS the
black-hole criterion leads again to $\delta
L\gtrsim l_\text{Pl}$ if we assume $s\sim L$ and argue that the quantum
mechanical uncertainty for a material particle scales as $\delta L
\gtrsim \sqrt{L/M}$.  This latter scaling is based on a semi-classical
picture \cite{Amelino-Camelia94} with an initial width of a
wave-package leading to a minimal 
width in momentum space, that is interpreted as particles
spreading out with a corresponding momentum distribution, giving a
correspondingly larger uncertainty for the position measurement at a
later time $T$.  The argument can be made more rigorous by minimizing
the quantum-mechanically calculated expectation value $\langle\delta
x(0)\delta x(t)\rangle$ of a particle by minimizing over its mass
\cite{Calmet04}.  One also recognizes in $\delta L
\gtrsim \sqrt{L/M}$ the standard quantum limit (SQL), and in particular
for $M=N\omega$ for a device dominated by the mass of $N$ photons a
scaling with $1/\sqrt{N}$. }

{\subsection{Quantum gravity theories and phenomenological models}\label{sec.rainbow}
For most microscopic theories of quantum gravity it is difficult to
extract bounds on minimal uncertainties of lengths.  In
\cite{hossenfelder_minimal_2013}, a generalized uncertainty principle (GUP) of the
form $\delta x ^\nu \delta p^\nu\gtrsim 1+l_sE$ is given as a
prediction of string theory, as well as a space-time uncertainty
$\delta x \delta T \gtrsim l_s^2$, where $l_s$ is a (yet unknown)
string scale that might be of the order of $l_\text{Pl}$, and $E$ the
energy with which the string is tested. 
In \cite{vasileiou_planck-scale_2015} it was stated that
Lie-algebra non-commutative 
space-times with non-commuting position coordinates,
$[x_\alpha,x_\beta]=iR\indices{_{\alpha\beta}^\gamma}x_\gamma/m_\text{Pl}$,
lead to a $\delta T$ of the form $\delta
T\sim L^n E^m/m_\text{pl}^{1+m-n}$ where $m,n$ are some
model-dependent powers with $1+m-n>0$.  The lowest-order non-trivial
case $n=m=1$ that gives an energy dependence, 
corresponds to $\delta T\sim L E/m_\text{Pl}$.  Considering $T$ as the
travel time of a particle from source to 
detector, $\delta T$ implies an
uncertainty of the radar length.   Combining this $\delta
T$ with the standard contribution from the Heisenberg uncertainty
principle and minimizing over the energy gives a minimal length
uncertainty that can be written in the form
\begin{equation}
  \label{eq:gendL}
\delta L\gtrsim l_\text{Pl}^\alpha L^{1-\alpha}  
\end{equation}
with some real value $\alpha\in [0,1]$ \cite{vasileiou_planck-scale_2015}. }

{Given the mentioned difficulty to extract predictions of fluctuations
of positions or lengths from microscopic
quantum gravity theories, mostly phenomenological
GUPs have been used to
generalize lower bounds based on the standard uncertainty principle.
It
is clear from dimensional grounds that \eqref{eq:gendL} is the generic
form of a 
power law scaling with $l_\text{Pl}$ if only $l_\text{Pl}$ and $L$
exist as length scales.  Such a form is therefore also obtained in
many other phenomenological theories, notably models that assume
fluctuations on the scale of the Planck length and then ask how these
accumulate during the propagation of a light signal.  The simplest
case are random walk models, which lead to $\alpha=1/2$
\cite{diosi_minimum_1989,amelino-camelia_gravity-wave_1999}; $\alpha=2/3$
is known as the holographic model.  If one assumes a fluctuation
$\delta \lambda$ of the wavelength $\lambda$ of the light used to
measure distances with $\alpha=1/2$, $\delta\lambda\gtrsim
l_\text{Pl}(\lambda/l_\text{Pl})^{1/2}$, the fluctuations of the total
length are given in the random walk model by $\delta L \gtrsim
\delta\lambda(L/\lambda)^{1/2}=l_\text{Pl}^{1/2}L^{1/2}$, i.e.~the new
length-scale $\lambda$ drops out.  However, if the fluctuations
$\delta\lambda$ are added up coherently, i.e.~all with the same sign,
a much larger value 
results,
\begin{equation}\label{eq:dLcoh}
\delta L\gtrsim (l_\text{Pl}L)^{1/2}(L/\lambda)^{1/2}.  
\end{equation}
The choice of model has therefore important
implications for the falsifiability of the predicted minimal
fluctuations.  E.g.~in \cite{lieu_phase_2003} the coherence of Hubble-space telescope
images of distant galaxies was used to bound possible quantum
fluctuations of space-time from below.  No fluctuations were found, but the 
coherent addition of the fluctuations was
subsequently questioned \cite{ng_probing_2003}.}

{Modified commutation relations lead in general to a generalized 
uncertainty principle.  In as much as this implies a fluctuating speed
of light, Lorentz invariance can be violated, but need not (see
e.g.~the model of discrete space-time with modified commutation
relations without violation of Lorentz invariance 
due to Snyder in 1947 \cite{snyder_quantized_1947}). In the same
way, the (deterministic) dispersion relation of e.m.~waves can be modified; such 
theories have become known as ``rainbow gravity''.  This class of theories contains  
doubly (or deformed) special relativity (DSR), with a kappa-deformed
Poincar\'e group
\cite{PhysRevD.67.044017,amelino-camelia_relativity_2002,liberati_interpreting_2005,sotiriou_quantum_2009,visser_lorentz_2009,weinfurtner_cosmological_2009}. DSR
is 
based on the idea that not only  
the speed of light is independent of the reference-frame, but also the
small length-scale $l_\text{QG}$ on which quantum-gravity effects
become important, 
identified typically with the  
Planck-length. 
DSR has recently been elaborated further into
"relative locality" 
\cite{PhysRevD.84.084010}, a
theory that emphasizes the importance of phase-space and suggests that 
momentum-space might be curved, which would imply non-linear conservation
laws of energy and momentum, and a relativity of ``locality''.  Another
formulation of DSR considered an energy-dependence of space-time
\cite{PhysRevLett.88.190403,PhysRevD.67.044017}. Earlier theories also
proposed a time-dependent speed of light as solution to cosmological
problems \cite{moffat_superluminary_1993,albrecht_time_1999}.}

{In \cite{PhysRevD.73.045020,PhysRevD.74.085017} it was proposed that
a non-linear dispersion relation might arise from averaging a
quantum-fluctuating metric over a relevant length scale of a test
particle.  Considering a ``measurement process'' in 
relativistic rather than quantum terms, it was suggested that 
the metric relevant for a measurement process of the momentum $p_\alpha$ of a
particle with energy 
$E$ is the ``classical'' metric of GR plus an averaged perturbation of
quantum-gravitational origin, assumed non-vanishing when averaging over the de
Broglie wavelength $\lambda=1/E$  of a deeply relativistic
particle,  thus introducing an extra 
energy-dependence into the (inverse) dispersion relation
$p_\alpha(E)$.  }


{In \cite{amelino-camelia_distance_1997} a modified dispersion relation
was found in the context of
a non-critical-string approach to quantum gravity. It leads to a
minimal total uncertainty of a length measurement based 
on the propagation of massless particles
\begin{equation}
  \label{eq:mindl}
  \delta L\gtrsim \sqrt{\eta L l_\text{Pl}}+l_\text{Pl},
\end{equation}
where $\eta$ is a dimensionless parameter of order one, and clearly
the first term dominates for $L\ll l_\text{Pl}$, giving
\eqref{eq:gendL} with $\alpha=1/2$, but $\alpha=1$ for $L\simeq
l_{Pl}$.  Underlying \eqref{eq:mindl} is
an assumption about the form of a decoherence-term in the modified
quantum Liouville equation that arises from coupling matter to the
degrees of freedom of space-time fluctuations that scales as $E^2/m_\text{Pl}$
with Planck-mass $m_\text{Pl}$ and energy $E$ of a particle.  When
generalizing this to a scaling $E^n/m_\text{Pl}^{n-1}$, a dependence
\begin{equation}
  \label{eq:Amelino97}
  \delta L\gtrsim L^{1/n}l_\text{Pl}^{1-1/n} 
\end{equation}
 was predicted, which is again of the form \eqref{eq:gendL}.}

{In \cite{PhysRevD.72.044019}, it was argued that a
finite minimal uncertainty of time measurements is linked to the
perturbative approach to quantization, whereas in a non-perturbative
approach in principle infinite resolution could be achieved, as long
as particle energies are not bound from above (as might happen with a
modified dispersion relation). On the other hand, the
authors find a finite minimum resolution both in perturbative and
non-perturbative approaches, with a minimum length uncertainty
\begin{equation}
  \label{eq:Galan1}
  \delta
L\gtrsim l_\text{Pl},
\end{equation}
 whereas for large background times $\bar{T}$
 \begin{equation}
   \label{eq:Galan2}
   \delta L\gtrsim \sqrt{l_\text{Pl}c\bar{T}},
 \end{equation}
 as in the Wigner-Salecker case \cite{RevModPhys.29.255,PhysRev.109.571}.   
 In
  \cite{amelino-camelia_gravity-wave_1999},  
 other estimates of
  length fluctuations were discussed, one of them scaling as $\delta L
  \gtrsim (l_{\rm QG}\,c\,T)^{1/2}$, where $l_{\rm QG}$ is expected to
  be $l_{\rm 
    QG}\gtrsim l_{\rm Pl}$, which for $L=c T$ is again in line with
  \eqref{eq:gendL} with $\alpha=1/2$.}

 {\subsection{Comparison with our bounds}}
{
When trying to compare these previously found bounds with ours, the
first thing to keep in mind, is that our bounds are fundamentally 
for $\delta c/c$, not $\delta L/L$.  This is important as there is no
quantum mechanical operator for the speed of light, hence one cannot
apply directly the standard Heisenberg uncertainty principle.  Rather,
we resorted to q-pet, which gives generalized uncertainty relations
\cite{braunstein_generalized_1996}. Secondly, our bounds are based
directly on the light field itself, not the quantum mechanical
uncertainty in the position of a clock, an MRS point, or a
test-particle. We have furthermore the choice of the state of the
probe, notably it can be a multi-photon state, whereas previous
derivations typically considered single-particle uncertainty relations, with a
state that saturates 
Heisenberg's uncertainty relation.  Moreover, since the QCRB is
optimized over all possible measurements of the light field and has a
clear interpretation in terms of the minimal uncertainty of an
estimator of $c$, there are no conceptual issues with the meaning of
the measurement on very small length scales. Questions on how
fluctuations at smaller length-scale add up do not arise.  In
random-walk models one might wonder why one should add up fluctuations
of the wavelength, as no measurements are made at that length
scale. In the q-pet approach, measurements on the length-scale of the
wavelength are included just as any other measurement of the light
field, and the uncertainty is the one of the best possible estimator
of $c$, rather than fluctuations of  a measured observable (whose
existence at a very small length scale might be questionable; this issue
was indeed  recognized as one of the most important ones in the field,
see Section 4.2.5 in \cite{hossenfelder_minimal_2013}).  }

{By using a light signal, another length-scale  comes into play, namely the wavelength
$\lambda$ of the light, as well as the propagation time, which
in a cavity can be much larger than the length of the
cavity. Depending on the quantum state used, $\lambda$ is still
present in the final result for the lower bound.}

{If we do translate our bounds for $\delta c/c$ into a bound for
fluctuations of length estimations $\delta L$ by assuming $\delta L=
T \delta c$ with fixed $T$, we see from \eqref{eq:dcopt} that  for the
optimal state we get 
back $\delta L\gtrsim l_\text{Pl}$ for $L=c\,T$, i.e.~this corresponds
to $\alpha=1$ in \eqref{eq:gendL}. However, for $T\gg
L/c$, one can get uncertainties much smaller than the Planck length,
a fact that was not reflected by previous bounds.  This insight
results naturally from the use of q-pet, where time appears as a resource
for more precise measurements,
in sync with experimentalists' habit to provide uncertainties per
square root of Hz for fair comparison. }

{For a coherent state in a lossless cavity, the lower bound of $\delta
L$ implied by \eqref{eq:dcoptF1}  reads $\delta L\gtrsim
l_\text{Pl}^{2/3}\lambda^{1/3}(L^2/(c\,T)^2)^{1/3}$. If $L=c \, T$,
this is as \eqref{eq:gendL} for $\alpha=2/3$, but with $L$ replaced by
$\lambda$. One might wonder if there is a deeper reason behind the
fact that a classical light signal reproduces the holographic model
concerning the scaling of the smallest $\delta L$ with
$l_\text{Pl}$. Compared to the coherently added up fluctuations
eq.\eqref{eq:dLcoh}, this is, in the optical domain, still a much
smaller value for any $L$ larger than about $10^{-12}$\,m.  }

{Given their fundamental measurement-based nature, our bounds can serve 
  for judging the falsifiability of quantum gravity
  theories and phenomenological models: Predictions of fluctuations
  {in a given space-time region 
  that are smaller than those given by 
  our bounds} can never be falsified
  through direct measurement {\em as a matter of principle} (subject
  to the made assumptions).  While the prefactors depending on
  $L,\lambda,T$ for the coherent 
  state matter, as a rule of thumb, predictions of fluctuations
  with $\alpha>2/3$ could not be measured with light in a coherent
  state, as the measurements own smallest possible uncertainty 
  $\propto l_\text{Pl}^{2/3}$ is larger. 
  Length uncertainties $\propto 
\sqrt{l_\text{Pl}L}$ of Wigner-Salecka-type theories as well as the
bound in \eqref{eq:mindl} are at least in principle falsifiable
with light in a coherent state.  
The fluctuations \eqref{eq:Amelino97} 
cannot be measured with light in a coherent state as soon as $n>3$, 
but they would be accessible at least in principle to ``quantum enhanced
measurements'' using the optimal quantum state of light.  However, it
is unlikely that an optimal state of light with a sufficiently large
photon number can ever be built, given the experimental difficulties
of producing superpositions of Fock states with even a few photons.
The
fluctuations predicted in 
\cite{Urban13} are well above our bounds for any cavity of realistic
size. }

{Several works discussed the possibility to measure fluctuations of
space-time created on the Planck-scale with gravitational wave
interferometers such as LIGO
\cite{amelino-camelia_gravity-wave_1999,Ng00,amelino-camelia_gravity-wave_2000}.
Bounds on $l_{\rm QG}$ were obtained from experimental data from
Caltech's 40\,m interferometer   \cite{abramovici_improved_1996}.   In
\cite{amelino-camelia_gravity-wave_2000} it was argued that  the
stated displacement noise level of that interferometer of order
$3\cdot 10^{-19}\,\text{m}/\sqrt{\text{Hz}}$ in the neighborhood of
450\,Hz already rules out length fluctuations of the interferometer
arms of order $l_\text{Pl}$ per Planck-time interval for the
random-walk accumulation of individual Planck-cell 
fluctuations to a total uncertainty. 
 {\cite{Amelino-Camelia98,Ellis00,Biller99,Kaaret99,Amelino-Camelia14,vasileiou_planck-scale_2015,Perlman}}  
attempted  to bound the supposed quantum fluctuations of space-time
using the broadening of light 
pulses from far-away astronomical sources, but so far the uncertainty
in the emission time of the 
light pulses as well as other sources of spreading the pulse are too large to say
much about quantum fluctuations of the metric \cite{Kaaret99}. }

\section{Concluding Discussion}
Our results 
imply  that {\em one should not think of quantum 
  fluctuations of space-time as
  existing independently of the measurement devices that probe them},  
  but rather as something that can only be defined in conjunction with
  them.  This is in 
  line with the modern theory of quantum measurement, where the
  possible measurement results do not only depend on the quantum
  system, but  also on the   quantum probe and its initial quantum
  state.  

{Accordingly, we find different lower bounds for $\delta c/c$ for the
optimal state and a coherent state. The former reproduces $\delta
L\gtrsim l_\text{Pl}$ when translated to the uncertainty  of a length and
assuming a measurement time $T\simeq L/c$,
whereas the latter is substantially enhanced and still depends on the
wavelength, scaling only as $l_\text{Pl}^{2/3}$.  Their derivation from
standard quantum optics
and GR is similar in nature to those of previous bounds
based on Gedanken-experiments (see \ref{sec.TE}) within QM and GR, but
provides a conceptual advance by the use of q-pet, which includes the
optimization over all possible measurements, and precise calculations
rather than orders of magnitude arguments. Simple scaling arguments
can be insufficient, as the discussions in the literature about how
fluctuations on small scales add up on long distances have shown.
Another example: in the Heisenberg microscope including gravity, one
might arrange the particle half way between light source and
detector.  In that case the acceleration due to the gravitational pull
will  average to zero and it is not clear why the quantum uncertainty
should be bounded from below by a gravitational effect --- not to talk
about questions of how the photon is supposed to be localized in
space-time, when only its wavelength is specified.  Such questions on how exactly
the measurement is done, and whether a different setup might not avoid
the limitations, do not arise in our q-pet approach.}

{Nevertheless, our bounds are of course subject to several
  (reasonable) restrictions as well: We
consider direct measurements of the propagation speed or phase speed of
an e.m.~wave.  Note, however, that the QCRB bounds the uncertainty for
any measurement and estimation scheme, as long as $c$ is
imprinted on the quantum state through the standard time evolution in quantum
optics with \eqref{eq:Hemf} as hamiltonian.  Ambiguities
arising from a proper definition of arrival time of the pulse pertain
to the level of different data analysis schemes and are fully covered
by the QCRB.  

We want to know the value of $c$
in a given region of  
space-time, and we assume a sufficiently rigid measurement apparatus
whose length 
remains unchanged when the photon number is increased. 
Apparatuses with finite
rigidity could deform under the influence of the gravity of the light
signal and the modification of Coulomb's law. For any
realistic material that
deformation should be negligible, however, compared to the one due to
the light pressure; this will be examined in more detail in another
publication \cite{Raetzel17}.  The
gravitative effect   
of the elastic energy was already shown in 
\cite{grishchuk_emission_1974} to be smaller than the one of the e.m.~field by a
factor $c_s/c$, where $c_s$ is the speed of sound in the cavity
walls.    We rely on the validity of 
quantum mechanics (more precisely quantum optics and q-pet) and GR in
semiclassical approximation (i.e.~$T^{\mu\nu}$ calculated as
q.m.~expectation value), and the validity of the linear dispersion
relation $\omega=c k$ for wave-lengths well 
above the quantum-gravity/Planck length.  For finding the optimal
state, we assume a maximum possible photon number in the state.}
We neglect  uncertainties in $c$ due to the expansion of the Universe
\cite{kopeikin_optical_2015}, non--inertial observers, local
gravitation potentials e.g.~from Earth or a (stochastic)
gravitational-wave (GW) background  
\cite{Abbott}, and quantum fluctuations of the mirror
positions.  In the quantum foam picture, also the latter should depend
on the way they are measured, but in any case can only lead to reduced
precision. 
The GW background at optical frequencies is expected to be extremely small,
but might dominate at frequencies around 100-1000\,Hz, where a large
number of gravitational sources is expected to exist, see  
\cite{Sorge}. However, to cavities much shorter than the GW wavelength 
(300-3000\,km for the above frequencies), the modified metric due to
the GW appears as 
uniform, and the GW effect can hence in principle be eliminated by a
cavity in free fall, in contrast to the GR effect of the light inside
the cavity.
More generally, any additional source of modification of the speed of
light may lead to tighter lower bounds on the uncertainty of $\delta
c/c$ than ours, but will not invalidate them. 

\acknowledgments
We thank Kostas Kokkotas, Nils Schopohl, Claus
Zimmermann, Julien Fra\"isse, Dennis R\"atzel and Friedrich Wilhelm
Hehl for discussions 
and a critical reading of the manuscript. 
The research of URF was supported by the NRF of 
Korea, grant Nos.~2014R1A2A2A01006535 and 2017R1A2A2A05001422. 

\appendix

\section{Single mode reduction of q-pet}
\label{singlemode}
We here prove that very generally for a given maximum amount of energy
the optimal quantum measurement of $c$ can be reduced to measuring a
single mode of fixed frequency put into the optimal state  $|\psi_{\rm
  opt}\rangle=(|0\rangle+|2n\rangle)/\sqrt{2}$.  
Starting point is the Hamiltonian $H$ for the e.m.~field, decomposed into
modes labelled by a mode-index $k$, consisting of 
wave-vector $\bk$ and polarization $\epsilon$. 
Then 
\begin{equation}
  \label{eq:Hemf}
  H=\sum_{k}\hbar \omega_k {n}_k=\hbar c\sum_{k}k {n}_k,
\end{equation}
with angular frequency $\omega_k=c k$ and $k=|\bk|$. The Hamiltonian
has the general form $H=c G$ with a Hermitian generator $G=\hbar
\sum_{k}k {n}_k$. It leads in a given state $\ket{\psi}$ and
propagation over total time $T$ to QFI \cite{braunstein_generalized_1996}
\begin{equation}
  \label{eq:Ic}
  I_{c}=4 \Delta G^2 T^2 \equiv 4(\langle G^2\rangle-\langle
  G\rangle^2)T^2 . 
\end{equation}
Let $G=\sum_i e_i \ketbra{i}$ be the spectral decomposition of $G$,
and $\ket{\psi}=\sum_{i=1}^N c_i\ket{i}$, where we assume that
$\ket{1}$ ($\ket{N}$) are the states of lowest (largest) energy
available. Then $\Delta
G^2=\sum_{i=1}^Np_i e_i^2-(\sum_{i=1}^Np_i 
e_i)^2$ with $p_i=|c_i|^2$ and $\sum_{i=1}^Np_i=1$. 
The Popoviciu inequality \cite{Popoviciu35} states $\Delta
G^2\le (e_N-e_1)^2/4$. It is saturated for $p_1=p_N=1/2,\,\,p_i=0$
else. 
The state $\ket{\psi}=(|1>+e^{i\varphi}|N>)/\sqrt{2}$ with an arbitrary phase
$\varphi$ saturates the inequality and thus   maximizes $I_c$.
 If $e_N$ or $e_1$ is
degenerate, only the total probability for the degenerate energy levels
is fixed to 1/2, and arbitrary linear combinations
in the degenerate subspace are allowed.  But the value of $\Delta G^2$
remains unchanged under such redistributions, and we may still choose
just two non-vanishing probabilities $p_1=p_N=1/2$. 
The derivation did not make use of the multi-mode structure
of the energy eigentstates. Hence, exactly the same minimal
uncertainty of $c$ can
be obtained by 
superposing the ground state of a single mode with a Fock state of
given maximum allowed energy as with an arbitrarily entangled multi-mode
state containing components of up to the same maximum energy. Setting $N=2n$
leads to the announced optimal single-mode state.

\section{Calculation of the metric perturbation}
\label{appendixmetric}
The vector  potential of the e.m.~field in the cavity in Coulomb gauge 
$\bA(\br,t)=\Upsilon q(t) 
\bv(\br)$, where $\Upsilon$ is a constant, $q(t)$ the time dependent
amplitude, and $\bv(\br)$ the mode function,
with components
  \begin{eqnarray}
    \label{eq:modes}
    v_x&=&{\cal N}e_x\cos k_xx\sin k_y y\sin k_z z,\nn
    v_y&=&{\cal N}e_y\sin k_xx\cos k_y y\sin k_z z,\nn
    v_z&=&{\cal N}e_z\sin k_xx\sin k_y y\cos k_z z.
  \end{eqnarray}
The polarization vector $\vec{e}=(e_x,e_y,e_z)$ is normalized to
length one, and is orthogonal to the $\bk$-vector
$\bk=(k_x,k_y,k_z)$, where $k_i=l_i\pi/L$, and
$l_i\in\mathbb{N}_0$, and at most one of three given $l_i$ can be zero. 
Therefore, there are two polarization directions
(transverse modes) for each $\bk$ vector, with the exception of cases
where one of the $l_i=0$, where only one polarization is possible. 
The request that the modes be orthonormal,
\begin{equation}
  \label{eq:norm}
  \int d^3r \bv_l(\br)\cdot\bv_{l'}(\br)=\delta_{l,l'}
\end{equation}
leads to ${\cal N}=\sqrt{8/V}$, and we can define the mode-volume
$V_l=V/8$.  Note that the index $l$ stands here for both the discrete
$\bk$ vector and the polarization direction (1,2). Finally, we choose
$\Upsilon=1/\sqrt{\epsilon_0}$, such that
\begin{eqnarray}
  \label{eq:AEH}
  \bA(\br,t)&=&\sum_l\frac{1}{\sqrt{\epsilon_0}}q_l(t)v_l(\br),\nn
  \bE(\br,t)&=-&\sum_l\frac{1}{\sqrt{\epsilon_0}}\dot{q}_l(t)v_l(\br),\nn
  \bH(\br,t)&=&\sum_l\frac{1}{\mu_0\sqrt{\epsilon_0}}q_l(t)\nabla\times
  v_l(\br).
\end{eqnarray}
After quantization, the amplitudes $q_l$ become the quadrature operators
of a harmonic oscillator,
$\hat q_l=\sqrt{\frac{\hbar}{2\Omega_l}}(\hat a_l+ \hat a_l^\dagger)$,
$\hat p_l=\frac{1}{i}\sqrt{\frac{\hbar
    \Omega_l}{2}}(\hat a_l-\hat a_l^\dagger)$, where $\Omega_l=|\bk_l|c$. 
In the semiclassical approach the 
energy-momentum tensor for a single mode with mode function $\bv$ is given by
the quantum mechanical expectation value \cite{Deser57,MTW},
\begin{eqnarray}
  \label{eq:tmn2}
  T^{00}&=&\frac{\hbar \Omega}{4}\left(-\left\langle
  (\hat a-\hat a^\dagger)^2\right\rangle
  \bv^2+\left\langle(\hat a+\hat a^\dagger)^2\right\rangle(\nabla\times\bv)^2/k^2\right),\nn
T^{0i}&=&\frac{i\hbar\Omega}{2k}\left(\left\langle \hat a^2\right\rangle-\left\langle \hat a^{\dagger
  2}\right\rangle\right)(\bv\times(\nabla\times\bv))_i,\nn 
T^{ij}&=&\frac{\hbar\Omega}{2}\left(\left\langle\left(\hat a-\hat a^\dagger
  \right)^2\right\rangle
  v_iv_j \right.
 \nn & & \left. -\left\langle\left(\hat a+\hat a^\dagger\right)^2\right\rangle(\nabla\times \bv)_i(\nabla\times \bv)_j /k^2\right) +T^{00}\delta_{ij}, \nn
\end{eqnarray}
where $k^2=\bk^2$, and we have used the symmetrized form
$(\hat q\hat p+\hat p \hat q)/2$ of the quantum mechanical operators for the
$T^{0i}$ components.

For a $(01M)$ 
mode, $l_x=0,l_y=1,l_z=M$ dictates $\vec{e}=(1,0,0)$ as unique possible polarization.  
For $M=1$, the frequency $\Omega_l=\sqrt{2}\pi c/L$, and
\bea
{\bm v} &=& \sqrt{\frac8V} \sin(\pi y/L ) \sin (\pi z /L) {\bm e}_x ,\nn 
\nabla\times{\bm v} &=& \sqrt{\frac8V}\frac\pi L \sin(\pi y/L ) \cos
(\pi z /L) {\bm e}_y \nn
& & - \sqrt{\frac8V}\frac\pi L \cos(\pi y/L ) \sin
(\pi z /L) {\bm e}_z .
\ea
For $\ket{\psi_{\rm opt}}$
with $n\gg 1$, and neglecting terms of order ${\cal O}(n^0)$ (all
other terms are of order $n$), we find that for the fundamental $(011)$ mode the 
only 
non-vanishing components of $T^{\mu\nu}$ can
be expressed in terms of four functions,
\begin{equation}  \label{eq:tmn3}
 T^{\mu\nu}=n\frac{\hbar\Omega_l}{V}t^{\mu\nu}
\end{equation}
with the dimensionless tensor components 
$t^{00}(\eta,\zeta)=f_1(\eta,\zeta)$,
$t^{11}(\eta,\zeta)=f_2(\eta,\zeta)$,
$t^{22}(\eta,\zeta)=f_3(\eta,\zeta)$,
$t^{33}(\eta,\zeta)=\tilde f_3(\eta,\zeta) = f_3(\zeta,\eta)$,
$t^{23}(\eta,\zeta)=t^{32}(\eta,\zeta)=f_4(\eta,\zeta)$, and
\begin{eqnarray}
  \label{eq:3fs}
f_1(\eta,\zeta)&=&2-\cos(2\eta)-\cos(2\zeta)\nn 
f_2(\eta,\zeta)&=&\cos(2\eta)+\cos(2\zeta)-2\cos(2\eta)\cos(2\zeta)\nn 
f_3(\eta,\zeta)&=&\frac{1}{2}(2-4\cos(2\zeta)+2\cos(2\zeta)\cos(2\eta))\nn
f_4(\eta,\zeta)&=&\sin(2\eta)\,\sin(2\zeta),
\end{eqnarray}
where we write $x,y,z$ in units of $L/\pi$, $\xi=x\pi/L$,
$\eta=y\pi/L$, $\zeta=z\pi/L$, and thus $\xi,\eta,\zeta\in 
[0,\pi]$. Outside the cavity $T^{\mu\nu}$ vanishes. 
For this state the field equations are solved with a time-independent
metric. The wave 
equation reduces to the Poisson equation, 
\bea
\Delta \bar h^{\mu\nu} = -16\pi \frac{G}{c^4} T^{\mu\nu}. 
\ea
The solution is obtained by integrating the
inhomogeneity $T^{\mu\nu}$ over with the Green's function of the
Poisson equation, i.e. 
\begin{eqnarray}
  \bar h^{\mu\nu}&=&\frac{4G}{c^4}\int\frac{T^{\mu\nu}(\bx')}{|\bx-\bx'|}d^3x'\nn
&=& {\mathcal P} \int_0^\pi\int_0^\pi
d\eta'd\zeta'I(\xi,\eta-\eta',\zeta-\zeta')t^{\mu\nu}(\eta',\zeta'),\nn \label{eq:hmnfin}  
\end{eqnarray}
where the parameter $\mathcal P$ is given by  
\begin{equation}
  \label{eq:Asuppl}
  {\mathcal P}=4\sqrt{2}\frac{n\hbar G}{\pi c^3L^2}=\frac{4\sqrt{2}n}{\pi}\kappa\mbox{,\quad\quad
  }\kappa=\left(\frac{l_{\rm 
        Pl}}{L}\right)^2 .
\end{equation}
The integral kernel reads
\begin{equation}
  \label{eq:I}
  I(\xi,\eta,\zeta)=\ln\left(\frac{\xi+\sqrt{\xi^2+\eta^2+\zeta^2} 
    }{\xi-\pi+\sqrt{(\xi-\pi)^2+\eta^2+\zeta^2
    }}\right).
\end{equation}
Numerical evaluation of the two remaining integrals in
Eq.\,(\ref{eq:hmnfin}) shows that they are of order one inside the
cavity, and decay rapidly outside, as is required by the boundary
conditions of a 
flat metric far from the cavity. 

For $\ket{\psi_{\rm coh}}$, we have to consider the full retarded solution
of the wave equation 
according to 
\begin{eqnarray}
  \bar
  h^{\mu\nu}&=&\frac{4G}{c^4}\int\frac{T^{\mu\nu}(t-|\bx-\bx'|/c,\bx')}{|\bx-\bx'|}
  d^3x'.  
\end{eqnarray}
For example, the $yz$ component reads
$\bar h^{yz}=
\bar h^{yz}_{\rm opt}  +\frac{4n\hbar G \Omega}{c^4} \int d\xi' d\eta' d\zeta' 
\frac{\sin[2\omega(t-|\bx-\bx'|/c)]  \sin (2\eta') \sin (2\zeta')}{ |\bx-\bx'| }$.
This metric element is thus the solution of  $\ket{\psi_{\rm opt}}$ \eqref{eq:hmnfin}
plus some retarded oscillation on top of it, which is of the same
order.  In the following we will therefore restrict our analysis to the
time-independent part given by  $\ket{\psi_{\rm opt}}$.

For the $(01M)$ 
mode, with $M>1$, $l_x=0,l_y=1,l_z=M$, the
general expressions for $T^{\mu\nu}$ are more complicated, but for
 $\ket{\psi_{\rm opt}}$ with $n\gg 2$, and in
the limit of $M\gg 1$, we have
$T^{00}=T^{33}=4n(\hbar\Omega/V)\sin^2\eta$,
$T^{11}=-T^{22}=4n(\hbar\Omega/V)\sin^2\eta\cos(2M\zeta)$.
Corrections are of
order $1/M$.  All other tensor elements of $T$ vanish to order
$M^0$.  The rapidly oscillating term $\cos(2M\zeta)$ in $T^{11}$,
$T^{22}$ leads to a rapid decay of
$\bar{h}^{11}$ and $\bar{h}^{22}$ as function of $M$. Numerics indicates that
the decay is roughly as $1/M$ for 
fixed $(\xi,\eta,\zeta$), including the factor $M$ that is gained due
to the prefactor $\Omega\propto M$ for large $M$.   This means that
for large $n$ and $M$, only 
$T^{00}=T^{33}$ are non-negligible, with
\begin{eqnarray}
 \bar{h}^{00}&=&\bar{h}^{33}\simeq {\mathcal P} M
  \tilde{h}(\xi,\eta,\zeta),\nn
\tilde{h}(\xi,\eta,\zeta)&:=& 4\int_0^\pi\int_0^\pi
d\eta'd\zeta'I(\xi,\eta-\eta',\zeta-\zeta')\sin^2\eta',\nn  \label{eq:h00Suppl}
\end{eqnarray}
where the dimensionless function $\tilde h(\xi,\eta,\zeta)$ is once more of
order 1 inside the cavity and falls off rapidly outside. So using a
higher mode has the effect of reducing the perturbation of the metric
essentially to two diagonal elements of the metric tensors, but
increases the perturbation by a factor equal to the mode-index
$M$. 

In all cases, the amplitude of the space-time perturbation
due to the e.m.~field in the cavity scales as
\begin{equation}
  \label{eq:gres}
  \overline{h}^{\mu\nu}\sim \left(\frac{l_{\rm Pl}}L\right)^2 \,n\,M,
\end{equation}
proportional to the number of photons $n$ in the cavity, the mode
index $M$, 
and the squared ratio $l_{\rm Pl}/L$ of Planck length $l_{\rm
  Pl}\simeq 1.62 \times 10^{-35}\,$m and size $L$ of the cavity. The
expression remains 
valid for the fundamental mode with $M=1$. 

We note that throughout our analysis we tacitly assume that the photon densities in the cavity are
small enough and the cavity sufficiently large, such that we stay well below the critical
(electric) field strength $E_c =
m_e^2c^3/(e\hbar)=1.3\times10^{18}$\,V/m,  where $m_e$ is the mass of
the electron, beyond which nonlinear corrections to Maxwellian
electrodynamics  
 due to polarization of the quantum vacuum become important
\cite{EulerHeisenberg}. 
This condition may be translated into a minimal
cavity size $L$ using an energy density $\ord (\hbar c n M/L^4)$ and a
critical energy density $\ord(\epsilon_0 E_c^2)$. We obtain that  
$L\gg (\hbar^{3/4} e^{1/2} \epsilon_0^{-1/4}m_e^{-1}c^{-5/4}) (n M)^{1/4}=
(2.1\times 10^{-13}\,$m)$(n M)^{1/4}$ for linear
electrodynamics in the cavity to hold. For the two types of cavities
considered 
and all combinations of $n_{\rm opt}$ 
and $M$, the lower bound on $L$ is satisfied by the
cavity sizes considered.

From $h_{\mu\nu}$ we now
calculate a local measure of the modification of the coordinate speed of light
defined through the geodesics of the modified metric.

\begin{figure}[t]
\centering
\includegraphics[width=\columnwidth]{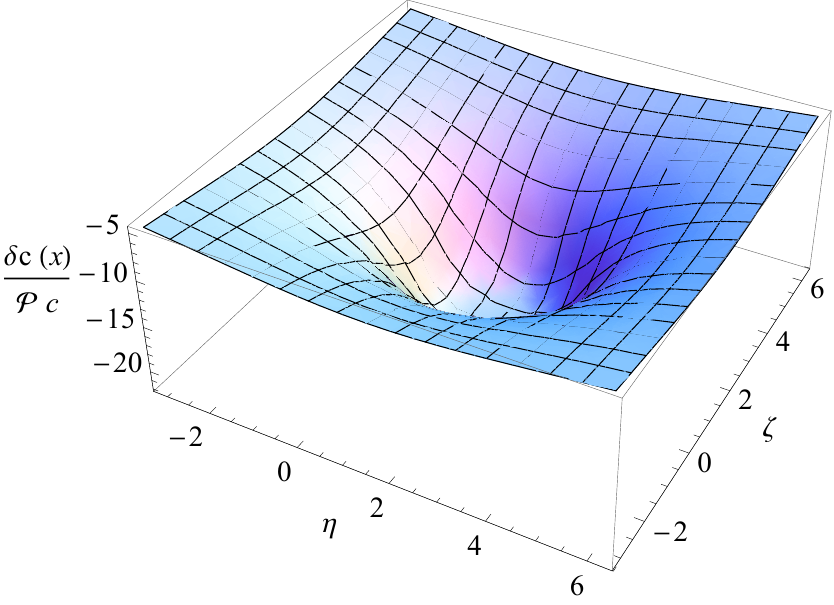}
\caption{Relative change of the local coordinate speed of light  in
  x-direction as function of dimensionless coordinates 
  $\eta,\zeta$ at $\xi=1.5$ in units of ${\mathcal P}=(4n/\pi)\kappa$ with
  $\kappa=(l_{\rm Pl}/L)^2$ (see Eq.\,(\ref{eq:Asuppl})) for the (011) mode. The
  cavity extends from 0 to $\pi$ in these units.
}
\label{fig.dc}
\end{figure}

A finite $h_{\mu\nu}$ leads to a new line element 
\begin{equation}
  \label{eq:ds2n}  
  ds^2= - (1- h_{00}) c^2 dt^2 + (1+h_{ii}) (dx^i)^2 
  +2h_{23}dy dz 
\end{equation}
where the metric elements are, for the (011) mode,
\bea
h_{00} &=& \frac12 {\mathcal P}(g_1+ g_2+g_3+\tilde g_3), \nn 
h_{11}&=&\frac12
{\mathcal P}(g_1+g_2-g_3-\tilde g_3),\nn
\quad h_{22}&=&\frac12{\mathcal P}(g_1-g_2 +g_3-\tilde g_3), \nn
\quad h_{33}&=&\frac12{\mathcal P}(g_1-g_2 +\tilde g_3-g_3), 
\quad h_{23}={\mathcal P}g_4,\label{eq:hij}
\ea
with the definitions, cf. \eqref{eq:hmnfin},
\bea\label{defgi}
g_i &=& \int_0^\pi\int_0^\pi
d\eta'd\zeta'I(\xi,\eta-\eta',\zeta-\zeta')f_i (\eta',\zeta') ,\nn 
\tilde g_i &=& \int_0^\pi\int_0^\pi
d\eta'd\zeta'I(\xi,\eta-\eta',\zeta-\zeta') \tilde f_i (\eta',\zeta') .
\ea
For the $(01M)$ mode we have $h_{00} =h_{33}$ with $h_{00}$ given by
(\ref{eq:hij})
 whereas $h_{\mu\nu}$ vanishes for all other
values of $\mu$,$\nu$. 
The light ray trajectories are determined through the geodesic
condition $ds^2=0$. The 
speed of light in $x^1$-direction (meaning all other $dx^j=0,\, j \neq
1$, i.e.~locally straight paths along $x^1=x$) is then $c(x)=c
\sqrt{(1-h_{00})/(1+h_{11})}$, and correspondingly for the other directions.
The relative change of the coordinate speed of light in $x^i$-direction
then reads, for the (011) mode with $n\gg 1$,
\begin{eqnarray}
  \label{eq:dc2suppl}
  \delta c(x)/c = -\frac12 (h_{00} + h_{11}) =
  -\frac12{\mathcal P}(g_1+g_2), 
\nn
  \delta c (y)/c = - \frac12 (h_{00} + h_{22}) =-\frac12 {\mathcal P} (g_1+g_3),
\nn   
  \delta c (z)/c =- \frac12 (h_{00} + h_{33}) = -\frac12 {\mathcal P} (g_1+
  \tilde g_3).
\end{eqnarray}
For the $(01M)$ mode with $n,M\gg 1$, 
\bea
\delta c(x)/c &=&\delta c(y)/c 
=-\frac12 h_{00} = -\frac{{\mathcal P}M}4(g_1+g_2+g_3+\tilde{g}_3),\nn
\delta c(z)/c&=&2\delta c(x)/c,
\eea 
where the equalities in terms of the $g_i,\tilde g_i$ are for 
$\ket{\psi_{\rm opt}}$.

In Fig.\,\ref{fig.dc},  we plot the relative change of the coordinate speed of
light in $x-$direction for the (011) mode. We see that up to position
dependent functions of order 1 the 
relative change of speed of light is given by Eq.\,(5) in the main text. 
Very
similar plots are obtained for other directions.


\bibliography{qcrmetCQG_final}





\end{document}